\documentclass[%
 reprint,
 amsmath,amssymb,
 aps,
]{revtex4-1}

\usepackage{graphicx}
\usepackage{dcolumn}
\usepackage{bm}

\usepackage{float}
\usepackage{subfigure}

\begin{document}

\preprint{APS/123-QED}

\title{Observations of highly localized oscillons with multiple crests and troughs}%

\author{{Xiaochen LI}}
\author{{Dali XU}}
\author{{Shijun LIAO}}
\email{sjliao@sjtu.edu.cn}
\affiliation{State Key laboratory of Ocean Engineering, School of Naval Architecture, Ocean and Civil Engineering, Shanghai Jiao Tong University, Shanghai 200240, China}%

\date{\today}

\begin{abstract}
Two types of stable, highly localized Faraday's resonant standing waves with multiple crests and troughs are observed in the alcoholic solution partly filled in a Hele-Shaw cell vertically oscillated with a single frequency.   These oscillons have never been reported.  Systematical experiments are performed to investigate the properties of these ocsillons.   Especially, some experiments are performed, for the first time, to indicate the high localization of the oscillons, which suggest that these  oscillons may be regarded as a combination of the two elementary oscillons discovered by Rajchenbach et al. (Physical Review Letters, {\bf 107}, 2011), for instance, (2,3) = (1,1) + (1,2), where ($m,n$) denotes oscillon with $m$ crests and $n$ troughs.  So, our experiments also reveal an interesting, elegant ``arithmetic'' of these oscillons.       
\begin{description}
\item[PACS numbers]47.54.-r, 47.35.Fg
\end{description}
\end{abstract}

\pacs{Valid PACS appear here}
\maketitle


Localized Faraday's resonant standing waves sloshing between the wall have been observed by Wu \cite{Wu1984} in the vertically oscillated water partly filled in a 3D trough. The measurements of the corresponding velocity field have been obtained by PIV and an extra streaming velocity has been found in \cite{Gordillo2014}. Recently, two new standing solitary waves with odd and even symmetries have been found in a Hele-Shaw cell partly filled with the distilled water \cite{Rajchenbach2011} when the cell is under vertically periodic vibrations. Since the distance between the walls of the Hele-Shaw cell is quite samll, there is no sloshing property for the two waves.  Kinds of other excitations in a layer of suspension have been observed as well in experiments. Both of harmonic/subharmonic oscillons and propagating solitons have been reported \cite{Lioubashevski1996, Arbell2000, Rajchenbach2013, Umbanhowar1996}. The transition from the oscillons to the propagating solitons has been investigated as well \cite{Lioubashevski1999}.  The mobility of the oscillon can be affected by the amount of protein in the suspension, and waves become well-organized by increasing the concentration of protein \cite{Shats2012}.  Instead, the ordered structure of the waves will break down if the amplitude of the vibration grows \cite{Xia2012}.  Interesting walkers and droplets have been observed as well in a similar situation when the acceleration reaches a critical threshold \cite{Couder2005}.  In this letter, we report the observation of some standing oscillons with multiple crests and troughs in a Hele-Shaw cell partly filled with the alcoholic solution. These oscillons have been never reported.  They are highly localized, but can be regarded as a combination of the two elementary oscillons (with odd and even symmetry) discovered by Rajchenbach et al. in \cite{Rajchenbach2011}. 

In our experiment, a thin layer of alcoholic solution in a Hele-Shaw cell is considered. The cell is made of polymethyl methacrylate (PMMA) and is 30 cm long and 1.7 mm breadth (the same size as \cite{Rajchenbach2011}). It is mounted on a shaker vertically oscillated with a displacement 
$z = A \sin{(2\pi f t)}$, where  the constants $A$ and $f$ denote the amplitude and frequency of the oscillation.  The acceleration amplitude is registered using an oscilloscope with an amplifier which is referenced externally to the input shaker signal.  The temperature is confined at about $20^{\circ}$.  Considering  the  volatility, the alcoholic solution is changed every ten minutes so as to guarantee the constant concentration.  Using a high-speed camera (250fps) positioned perpendicular to the cell,  the images are acquired once the oscillons have stable and distinct forms with a typical lifetime of about $1000$ oscillation periods. 

\begin{figure}[h]
\begin{subfigure}
{\rotatebox{0}{\scalebox{0.11}{\includegraphics{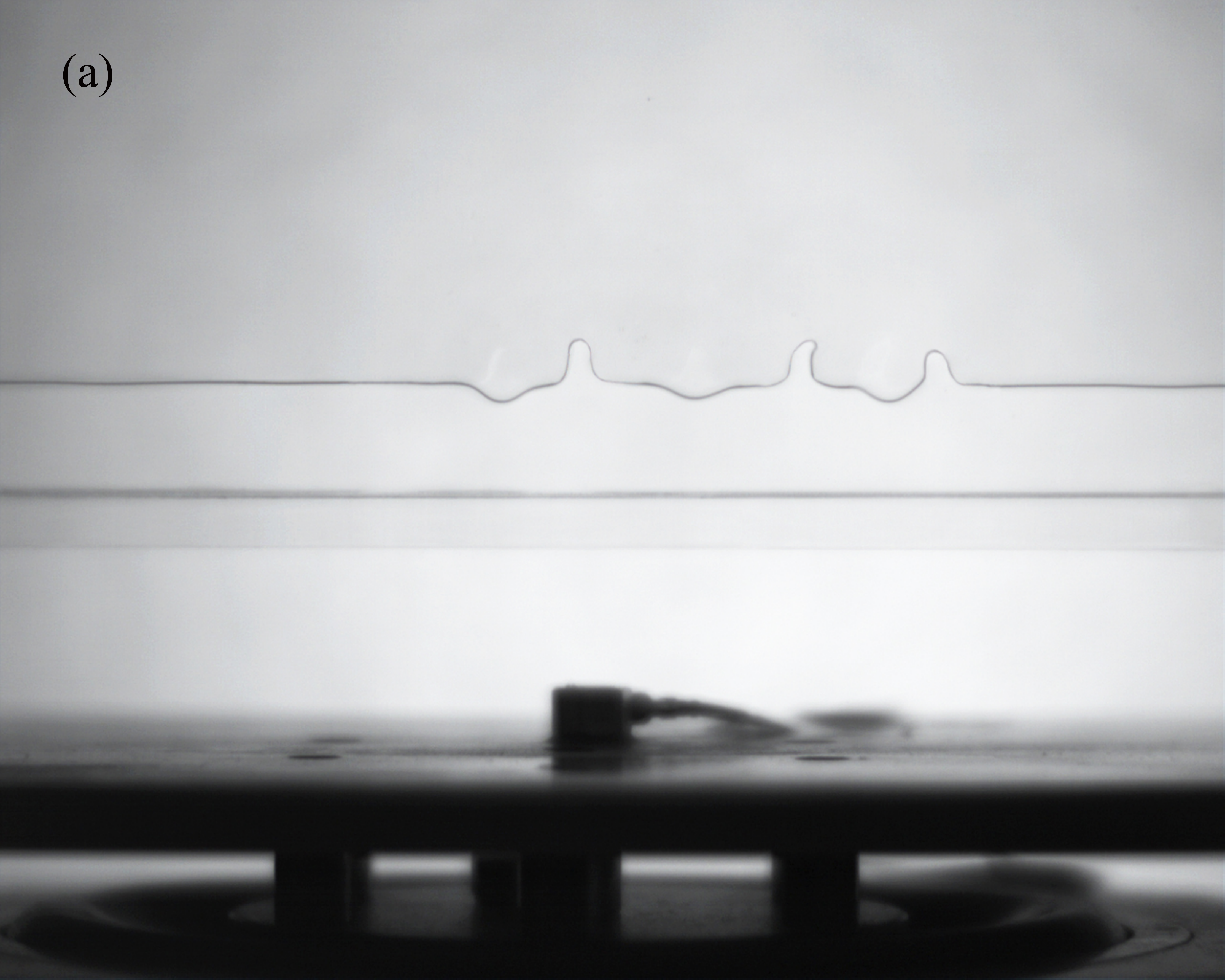}}}}
\end{subfigure}
\begin{subfigure}
{\rotatebox{0}{\scalebox{0.11}{\includegraphics{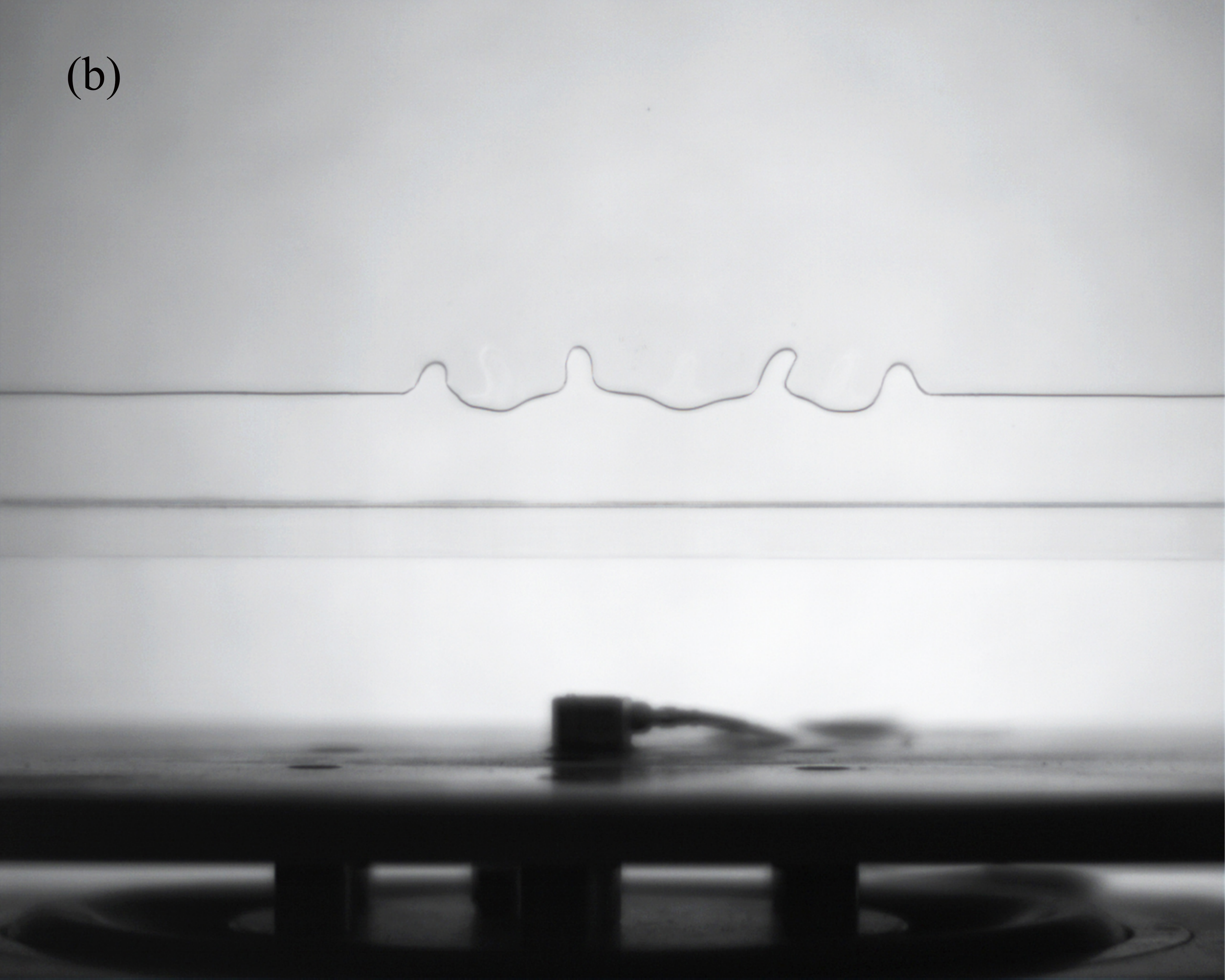}}}}
\end{subfigure}
\caption{\label{fig:epsart}  {\bf The oscillons (3,3) and (4,3)}.  The new oscillons observed  in $15\%$ alcoholic solution with fluid depth of 2 cm, oscillation frequency of  18 Hz and acceleration amplitude of 20.503 $\mathrm{m/s^2}$. {\bf (a)} and {\bf (b)} represent the typical images of the oscillons (3,3) and (4,3), respectively.  The corresponding supplementary movies are available online.}
\end{figure}

First, the experiments were  performed in a $15\%$ alcoholic solution with fluid depth of 2 cm, oscillation frequency of 18 Hz, and acceleration amplitude of 20.503 $\mathrm{m/s^{2}}$.  We disturbed the free surface by an extra probe as the initial, local excitation.  Dependent upon the way we disturbed the free surface,  stable subharmonic standing resonant waves with different shapes are observed.   When the disturbed region of free surface is short (but strong enough to excite the resonance), the same oscillons with the even and odd symmetry reported in \cite{Rajchenbach2011} are observed.  It is very interesting that new oscillons with multiple crests and troughs are observed when the disturbed region of free surface is enlarged, or an additional disturbance is given near the free surface of an observed oscillon.  They are highly localized, as shown in Fig.~1, and have stable and distinct forms with a typical lifetime of $1000$ oscillation periods.  Let $(m,n)$ denote  such an oscillon  with $m$ crests and $n$ troughs at time $t$. The oscillon  $(m,n)$ at $t$  becomes $(n,m)$ at $t+T/2$, namely that the $m$ crests and $n$ troughs at $t$ become the corresponding $m$ troughs and $n$ crests after a half period without changing the location.  So,  $(m,n)$ and $(n,m)$ denote the same oscillon.  These oscillons can be divided into two types.  One corresponds to $m=n$, the other to  $|m-n|=1$.  Note that $(1,1)$ and $(1,2)$ correspond to the elementary oscillons with the odd and even symmetry reported in \cite{Rajchenbach2011}.  Fig.~1a and b present the typical images of the oscillons (3,3) and (4,3),  when their crests reach the maximum positions.   At the acceleration amplitude  of $20.503$ $\mathrm{m/s^{2}}$,  twelve oscillons, say, (1,1), (1,2), (2,2), (2,3), (3,3), (3,4), (4,4), (4,5), (5,5), (5,6), (6,6) and (6,7),  are observed.   It seems that localized oscillons with more crests and troughs can be observed, if the cell is long enough.       

Secondly,  using $15\%$ alcoholic solution, we fixed the frequency at 18 Hz and the acceleration amplitude at 20.503 $\mathrm{m/s^{2}}$, respectively, but changed the fluid depth from 1 to 5 cm so as to investigate the influence of fluid depth on the ocsillons.   Without loss of generality, the (2,2) oscillon was studied.  The  adjacent crest-to-trough height was recorded when the crests and troughs approached their corresponding highest and lowest positions, respectively.    Two values of the adjacent crest-to-trough height were  recorded when the two crests of the oscillon (2,2) approached their highest position.   After a half period, another two values  were recorded.  Averaging these four values,  the so-called averaged wave height of the oscillon (2,2) was obtained.   Considering the uncertainty of the initial disturbance of free surface, five independent experiments were done (in fact, all experimental results reported in this letter were obtained in this way), so that five values of the averaged wave height of (2,2) are given.  The five wave heights have an averaged value 10.45$\pm$0.48 mm (average$\pm$standard  deviation).  Thus, it is reasonable to use the averaged value 10.45 mm as the averaged wave height of the oscillon (2,2).   It is found that the averaged wave heights in different fluid depths are almost the same, as shown in Fig~2a.  This suggests that the averaged wave height of the oscillon (2,2) is almost independent of fluid depth.  It is interesting that the oscillon is still localized and stable with the same wave height even in fluid depth of 1 cm, corresponding to a highly nonlinear dynamic system.   To the best of our knowledge, no theoretical result have been reported for this kind of oscillon.  It is found that fluid depth has almost no influence on other types of oscillons, too.  

To investigate the influence of the concentration of alcoholic solution, the same experiments were performed with the $10\%$ alcoholic solution at frequency of 18 Hz,  acceleration amplitude of 20.503 $\mathrm{m/s^{2}}$, and  fluid depth of 2 cm.  The same averaged wave height 10.45$\pm$0.11 mm of the (2,2) oscillon is observed.   This suggests that the concentrations of  alcoholic solution have almost no influence on the oscillons in general.  Thus, the  concentration of $15\%$ is used in all experiments reported below.   

\begin{figure}[h]
\begin{subfigure}
{\rotatebox{-90}{\scalebox{0.27}{\includegraphics{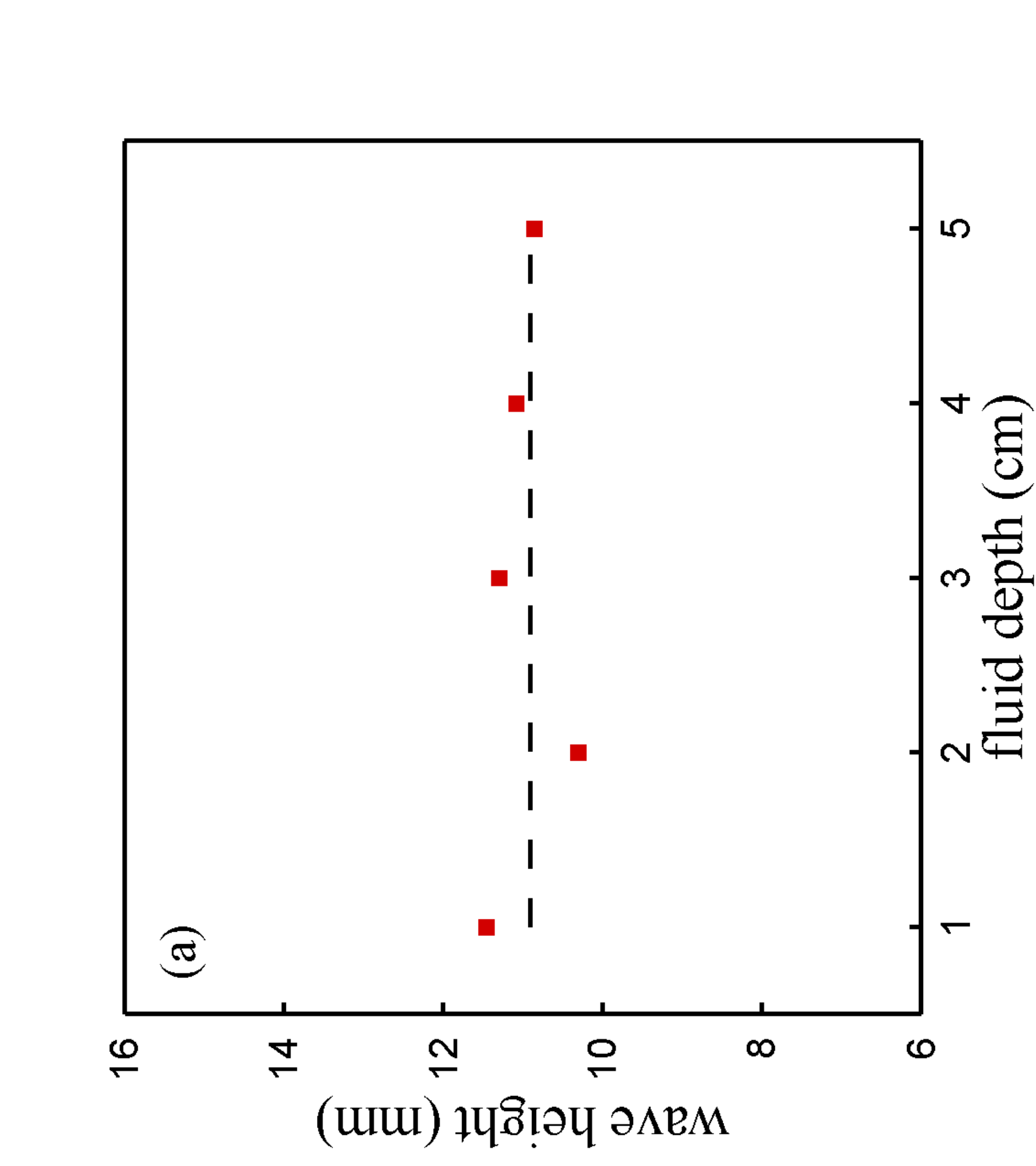}}}}
\end{subfigure}
\begin{subfigure}
{\rotatebox{-90}{\scalebox{0.27}{\includegraphics{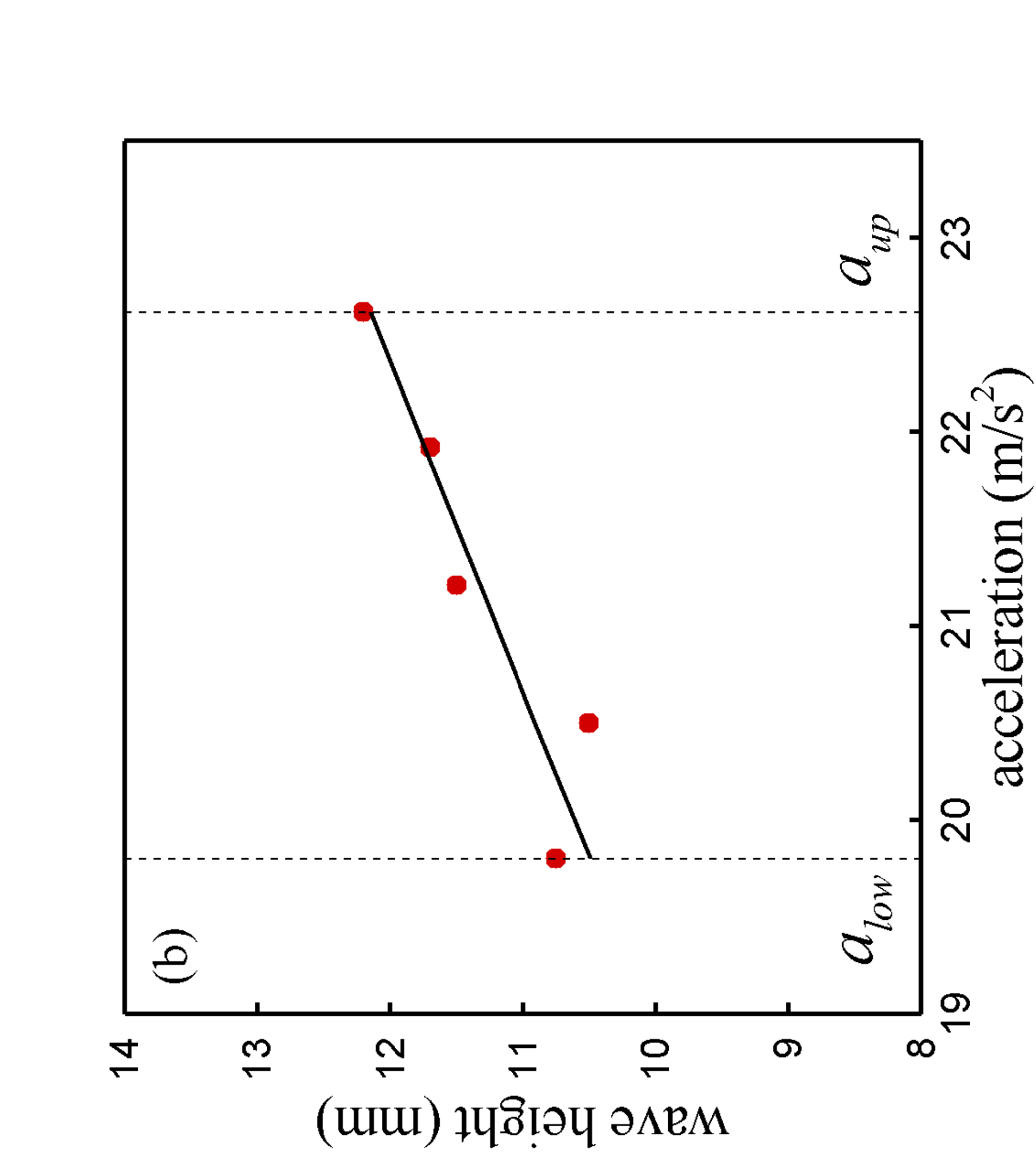}}}}
\end{subfigure}
\caption{\label{fig:epsart}  {\bf Averaged wave height versus fluid depth and acceleration amplitude}.  {\bf (a)}   Wave height of the (2,2) oscillon versus fluid depth,  in a $15\%$ alcoholic solution  at the frequency of 18 Hz and the acceleration amplitude of 20.503 $\mathrm{m/s^{2}}$.   The dots denote the experimental results, with a  dashed line representing  the averaged value 10.45 mm.  {\bf (b)} Wave height of the (2,2) oscillon versus acceleration amplitude, in a $15\%$ alcoholic solution at frequency of 18 Hz and fluid depth of  2 cm.  Dots denote the experimental results, with  a solid line for the optimal approximation.  $a_{low}$ and $a_{up}$ denote the lower and  up thresholds of acceleration amplitude for the existence of oscillons, respectively.  }
\end{figure}

To investigate the influence of the acceleration amplitude of oscillation on the oscillons, experiments were performed using  $15\%$ alcoholic solution at  frequency of 18 Hz and  fluid depth of  2 cm.   Without loss of generality, oscillon (2,2) was studied.  It is found that (2,2) oscillons are observed in a region of the acceleration  amplitude $[19.796, 22.624]$, and their averaged wave heights increase almost linearly, as shown in Fig.~2b.  It is found that there exist the lower threshold $a_{low}=19.796$ $\mathrm{m/s^{2}}$  and the upper  $a_{up}=22.624$ $\mathrm{m/s^{2}}$.  When the acceleration amplitude is below the lower threshold $a_{low}$, the oscillon has short lifetime and disappears after only a few periods of oscillations.   Instead,  when the acceleration amplitude is over the upper threshold $a_{up}$,  the oscillon is not localized and expanded into the whole cell so that the surface becomes wavy.  Note that the similar window of acceleration amplitude has been reported in \cite{Rajchenbach2011} for the elementary  oscillons (1,1) and (1,2), with a theoretical explanation.   It is found that  the  almost  same  lower  and up  thresholds existed  for other types of oscillons $(m,n)$.  Thus, our experiments indicate that such kind of window of the acceleration amplitude exists for the localized oscillon $(m,n)$ in general, with the linearly increase of averaged wave height with respect to the acceleration amplitude within this kind of existence-window.      

\begin{figure}[h]
\rotatebox{-90}{\scalebox{0.27}{\includegraphics{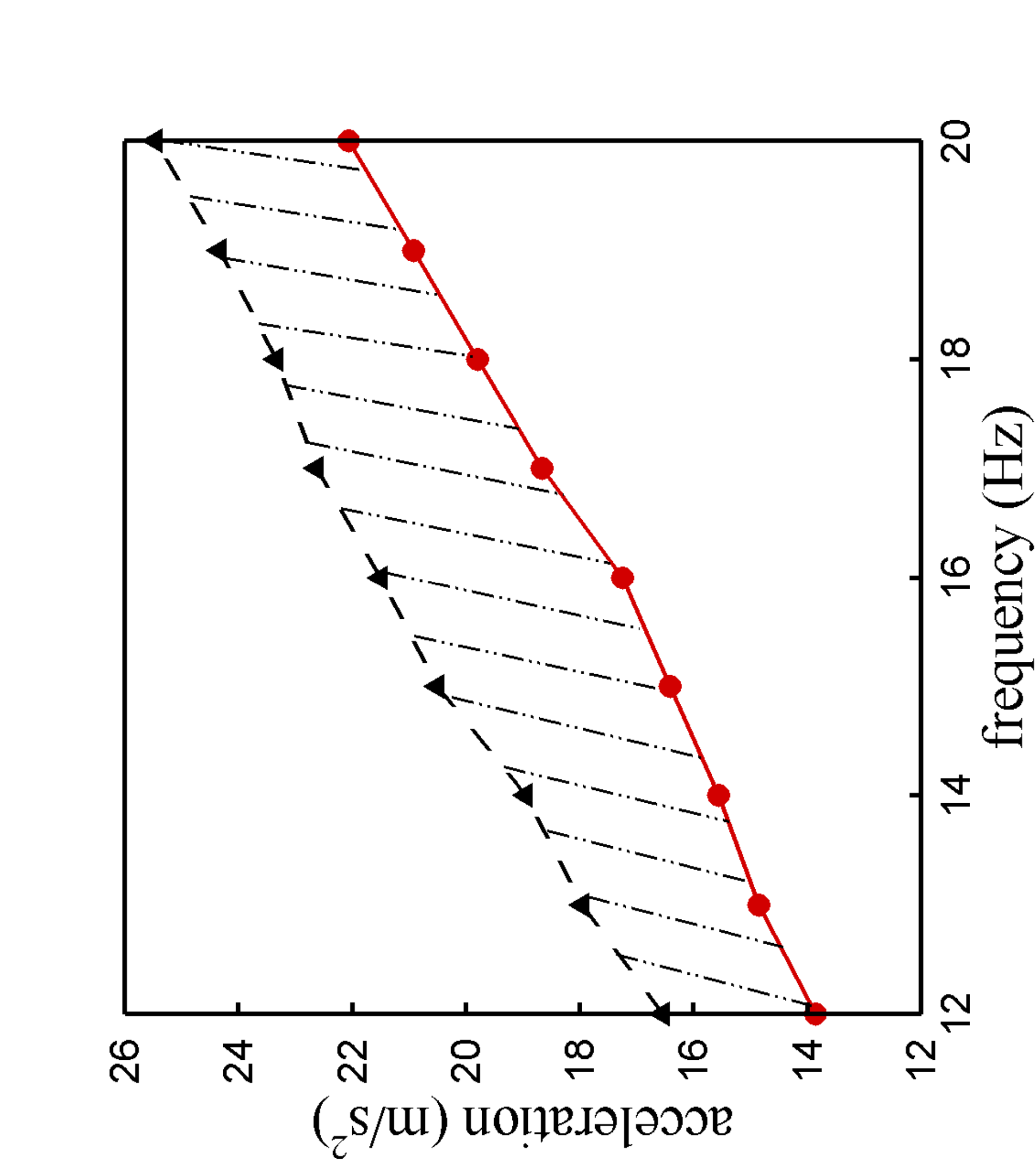}}}
\caption{\label{fig:epsart} {\bf Existence region of  oscillons}.  Experiments were performed in $15\%$ alcoholic solution with fluid depth of 2 cm.  Solid line with circles denotes the lower threshold of acceleration amplitude,  and the dashed line with triangles represents the upper one, respectively.}
\end{figure}
 
Do such kind of windows exist in general?  To answer this question, experiments were performed using $15\%$ alcoholic solution at the same fluid depth of 2 cm but different frequency of oscillation.  It is found that the similar window of acceleration amplitude  exists for frequency of oscillation from 12 Hz to 20 Hz, as shown in Fig.~3.  Both of the lower and up thresholds of the existence-window increas as the frequency enlarged, and various types of localized oscillons are observed within the windows.   It is found that oscillons have larger size at lower frequency, and some oscillons can not be observed below 12 Hz.  Besides, stable oscillons with long enough  lifetime can not be observed at frequency larger than 20 Hz.               

\begin{figure}[h]
\begin{subfigure}
{\rotatebox{-90}{\scalebox{0.27}{\includegraphics{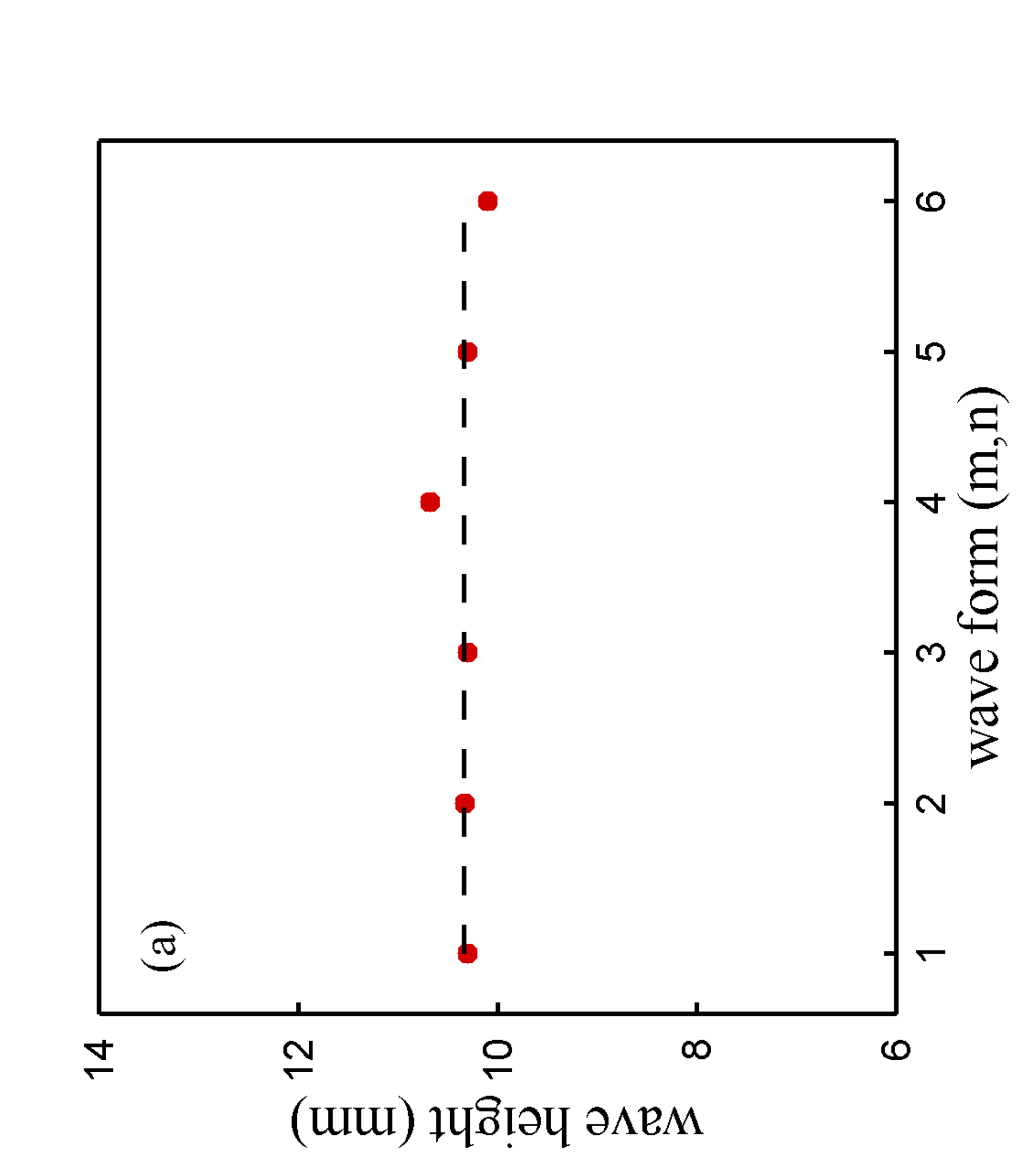}}}}
\end{subfigure}
\begin{subfigure}
{\rotatebox{-90}{\scalebox{0.27}{\includegraphics{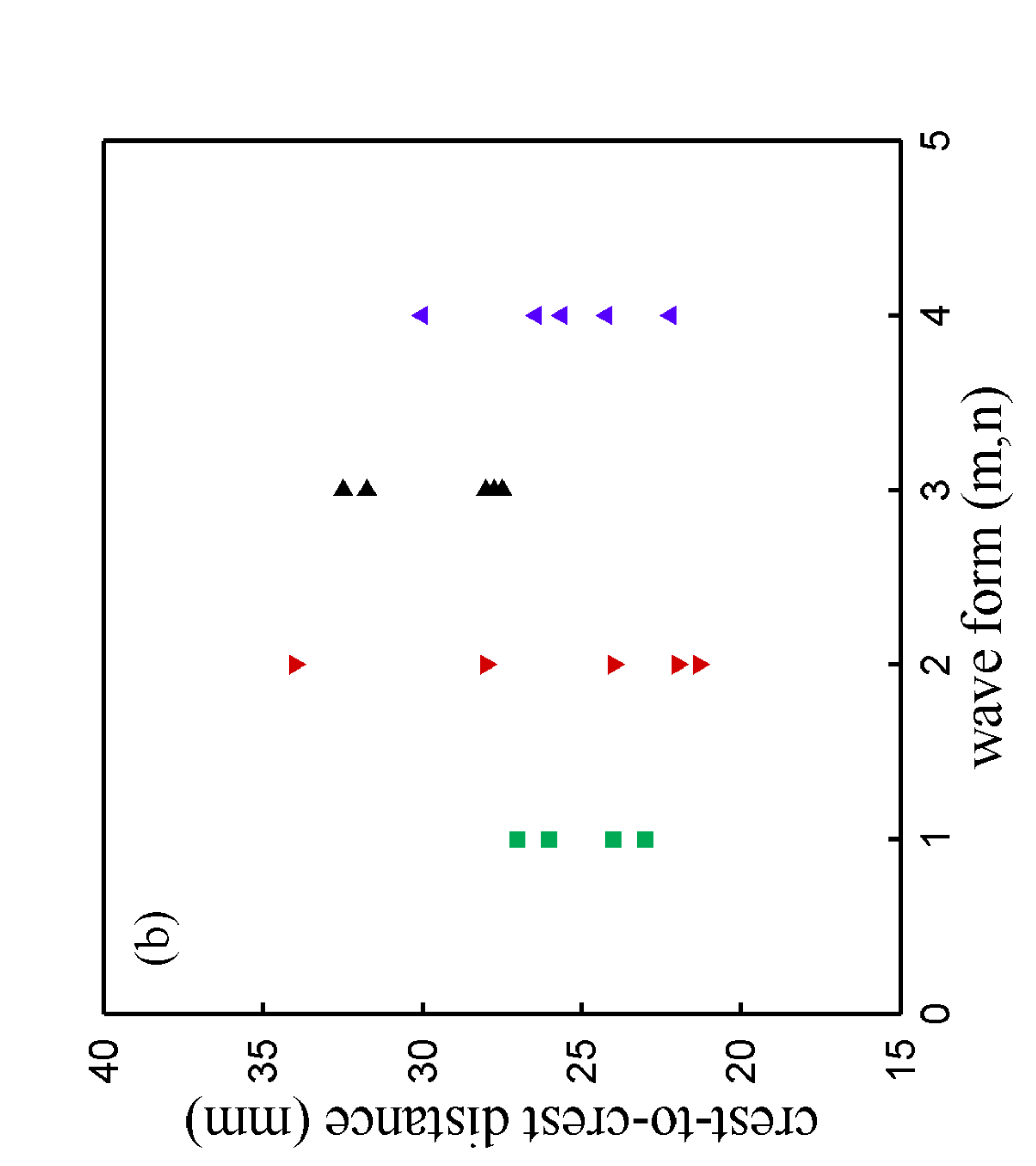}}}}
\end{subfigure}
\caption{\label{fig:epsart}  {\bf Wave height and crest-to-crest distance of oscillons}.   Experiments were performed in  $15\%$ alcoholic solution at frequency of 18 Hz,  acceleration amplitude of 20.503 $\mathrm{m/s^{2}}$, and  fluid depth of 2 cm.   {\bf (a)} The averaged wave height of different types of oscillons.  The integers  1 to 6 in the horizontal axis denote the oscillons of (1,1), (1,2), (2,2), (2,3), (3,3) and (3,4), respectively.  Dots denote the experimental results, with a dashed line representing their averaged value. {\bf (b)} The averaged crest-to-crest distance of different types of oscillons. The integers  1 to 4 in the horizontal axis denote the oscillons of (2,2), (2,3), (3,3) and (3,4), respectively.   Each symbol represents an experimental result.}
\end{figure}

\begin{figure}[h]
\begin{subfigure}
{\rotatebox{0}{\scalebox{0.11}{\includegraphics{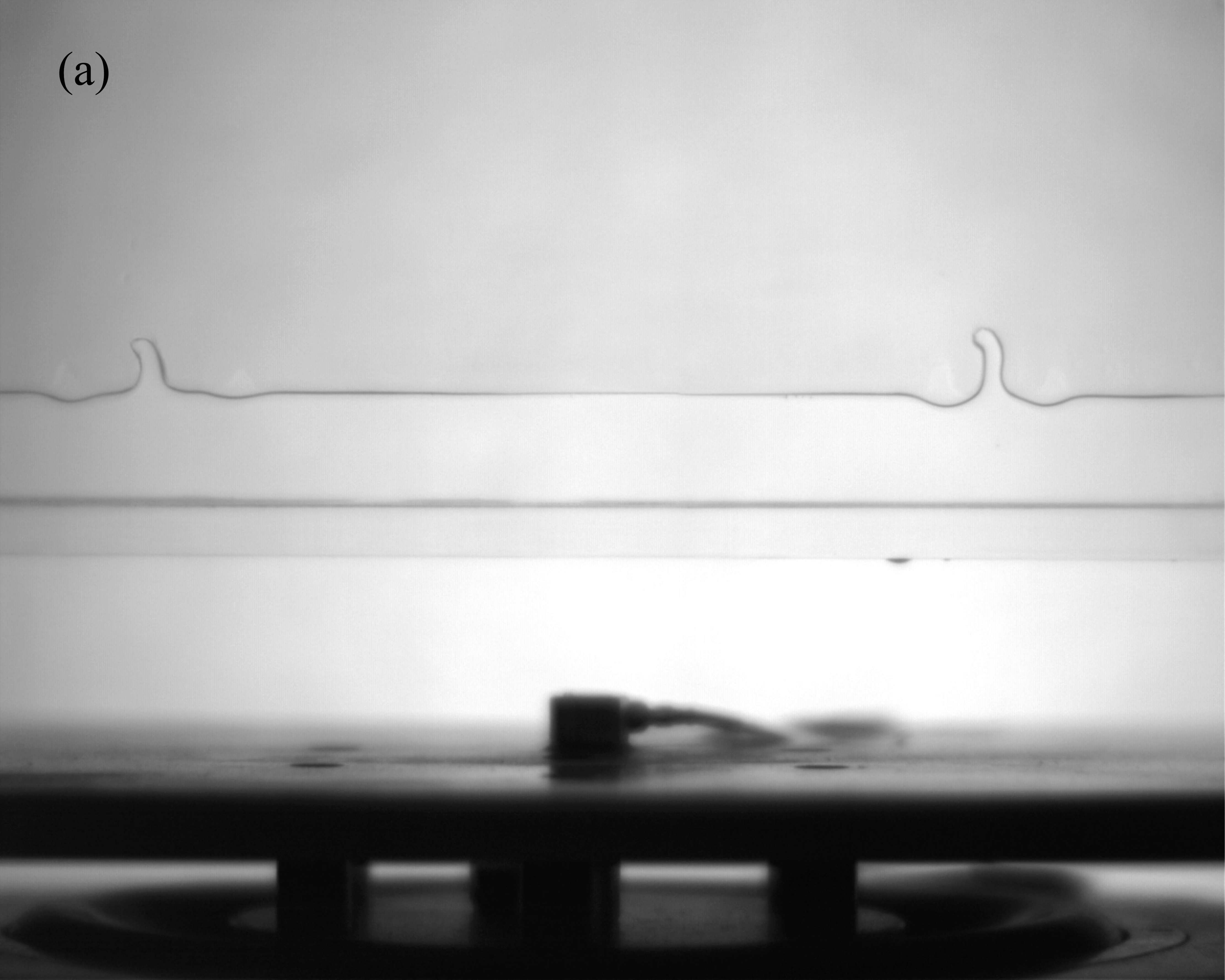}}}}
\end{subfigure}
\begin{subfigure}
{\rotatebox{0}{\scalebox{0.11}{\includegraphics{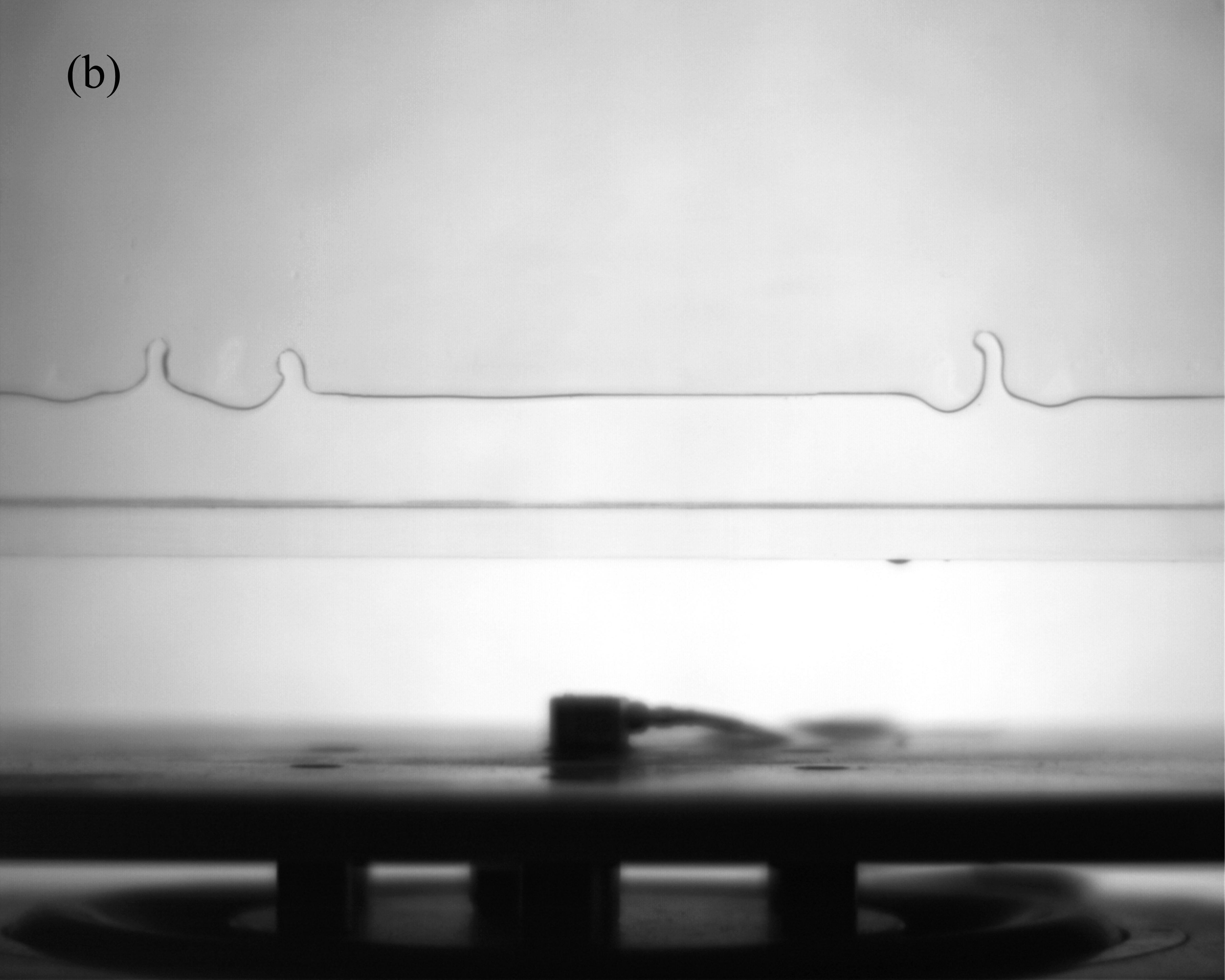}}}}
\end{subfigure}
\caption{\label{fig:epsart}  {\bf High localization of oscillons}.  Experiments were performed in $15\%$ alcoholic solution at frequency of 18 Hz,  acceleration amplitude of 20.503 $\mathrm{m/s^{2}}$ and fluid depth of 2 cm.  The right oscillon (1,2) keeps the same  in (a) and (b).  The left oscillon is first (1,2)  in (a)  but then becomes (2,2) in (b) by additional disturbance of free surface.}
\end{figure}

To compare the averaged wave height of different types of oscillons $(m,n)$, the experiments were performed with $15\%$ alcoholic solution at frequency of 18 Hz,  acceleration amplitude of 20.503 $\mathrm{m/s^{2}}$, and  fluid depth of 2 cm.   It is a little surprised that the averaged wave-heights of all types of observed oscillons are almost the same, say, 10.3$\pm$0.19 mm, as shown in Fig.~4a.  Especially, the oscillons (1,1) and (1,2)  have the same averaged wave height.  It suggests that the different types of oscillons have the same averaged wave height  for the fixed frequency and acceleration amplitude of oscillation at the same fluid depth. The adjacent crest-to-crest distance of these oscillons can be defined and measured in a similar way to give the so-called averaged crest-to-crest distance. It is found that  the averaged crest-to-crest distance of these oscillons are irregular, with the maximum gap of  12.7 mm and the maximum standard deviation of 20.2\%, as shown in Fig.~4b. The almost same wave-height but the irregular distribution of the averaged crest-to-crest distance strongly suggest that the observed oscillons are highly localized  so that they are a combination of the elementary  oscillons  (1,1) and/or (1,2).  For example,  the oscillon (2,2) can be regarded as the two (1,1) oscillons, i.e. $(2,2)=(1,1)+(1,1)$, which can be close enough since each (1,1) oscillon is highly localized.   Similarly, (2,3) can be regarded as a combination of one (1,1) oscillon and one (1,2) oscillon, i.e. $(2,3)=(1,1)+(1,2)$, which are close enough.  Since the elementary oscillons (1,1) and (1,2) have the same averaged wave height, as shown in Fig.4a, it is easy to explain why the more complicated oscillons like (2,2) and (2,3) have the same wave height.  However, since the distance between each (1,1) and/or (1,2) elementary oscillon can be different due to the  experimental uncertainty of the initial disturbance on free surface,  the so-called averaged crest-to-crest distance should be irregular with large gap and standard deviation.                                    

Some experiments were performed to support our above viewpoints, with $15\%$ alcoholic solution at frequency of 18 Hz,   acceleration amplitude of 20.503 $\mathrm{m/s^{2}}$, and  fluid depth of 2 cm.   At first, one stable (1,2) oscillon was excited at the right side of the cell.  Then,  the flat free surface at the left side of the cell was disturbed to excite another (1,2) oscillon, as shown in Fig.5a.  It is found that the right (1,2) oscillon keeps the same even after the left (1,2) oscillon is excited.  Besides, both of the left and right (1,2) oscillons have the almost same wave height and  shape.  After around $1000$ oscillation periods, additional disturbance was given to the left (1,2) oscillon, and a new (2,2) oscillon was excited. It is found that the right (1,2) oscillon keeps unchanged in the meanwhile.  These highly suggest that the two oscillons on the left and right side of the cell are independent of each other, say, highly localized.  Therefore, our experiments highly suggest that the observed, more complicated $(m,n)$ oscillons reported in this letter may be a combination of several  (1,1) and/or (1,2) elementary oscillons discovered in \cite{Rajchenbach2011}.  This viewpoint can well explain the experimental results reported above.                 


This work is partly supported by  National Natural Science Foundation (Approval No. 11272209), State Key Laboratory of Ocean Engineering (Approval No. GKZD010063) and the Lloyd's Register Foundation (LRF).  


\end{document}